\documentclass[11pt]{article}             
\usepackage{osidmod}                        
\usepackage{amsmath}
\usepackage{cite}
\usepackage{colortbl}

\definecolor{gray}{rgb}{0.7,0.7,0.7}

\title{Non-decomposable Quantum Dynamical Semigroups and Bound Entangled States}
\author{Fabio Benatti \\{\footnotesize\it Dipartimento di Fisica Teorica, Universit\`a di Trieste \& Istituto Nazionale di Fisica Nucleare, Sezione di Trieste, Trieste, Italy}\\[2ex]
Roberto Floreanini\\{\footnotesize\it Istituto Nazionale di Fisica Nucleare, Sezione di Trieste,
                     Trieste, Italy}\\[2ex]
Marco Piani \\{\footnotesize\it Dipartimento di Fisica Teorica, Universit\`a di Trieste \& Istituto Nazionale di Fisica Nucleare, Sezione di Trieste, Trieste, Italy}          }

\begin{document}

\maketitle
\begin{abstract}
We use open quantum system techniques to construct one-parameter
semigroups of positive maps and apply them to study the entanglement 
properties of a class of 16-dimensional density matrices,
representing states of a $4\times 4$ bipartite system.
\end{abstract}

\section{Introduction}

The recent developments in quantum information have turned entanglement 
into a concrete physical resource that is important to identify,
quantify and classify ({\it e.g.} see \cite{NC,NC2} and references therein).
The techniques devised to this end have spurred the novel use of 
some mathematical notions like positive linear maps \cite{CH1,Sto,ZB}
which had been somewhat neglected in quantum theory; in fact,    
fully consistent physical operations have to be described by
completely positive maps \cite{CH2,Tak,Kra}.

The need for complete positivity is due to the existence of
entangled states: entangled bipartite states,
if subjected to positive transformations acting on one
partner only, do not in general remain positive and thus
lose their probabilistic interpretation (for recent related work
on this issue, see \cite{BFBa,BFR3,BFR4,BFP1,BFPR}).
However, this very same fact renders positive maps useful
in detecting entanglement \cite{Per,HHH1,H1,HHH2,HHH0}.

The simplest instance of positive, but not completely positive map, is the
transposition: when acting on one partner only (partial transposition)
of a bipartite system
it detects all entangled states of
$2\times 2$ and $2\times 3$-dimensional systems.
In higher dimension, however, there are bipartite states
which remain positive under partial transposition and are 
nevertheless entangled.
Further, the kind of entanglement they contain cannot be
distilled by local quantum operations and classical communication 
(these and other properties are reviewed {\it e.g.} in \cite{HHH0}).

This undistillable entanglement is termed ``bound'' and
can be detected only by means of indecomposable positive maps, namely
maps that cannot be written as the sum of a
completely positive map and a completely positive map composed with
transposition.
It is thus of practical importance to provide
as many examples of indecomposable maps as possible, but
the task is rather difficult for, unlike completely positive maps, 
the structure of positive ones is still elusive \cite{CH1,Sto}.

In this work, we add to the phenomenology of positive maps \cite{CH3,Kye1,Kos}
a class of one-parameter semigroups \cite{BFP2} arising
from the theory of open quantum systems
\cite{Spo,AL,BP,BF0}; we discuss their decomposability and 
show how they can be used to detect bound entangled states.
In particular, we will be concerned with a class of $16\times16$
density matrices naturally arising from a $4\times4$ 
square lattice through the use of tensor products of Pauli matrices.

In Section 2, we shortly review some basic notions and results concerning
positivity, complete positivity and quantum dynamical semigroups; 
in Section 3, we focus on the above mentioned class of lattice
states and study which of them remain positive
under partial transposition (PPT); finally,
we present some results on the study of their
entanglement properties.
 
\section{Positive and completely positive maps}

We start with some basic facts about positive maps and entanglement in the
case of a finite $d$-dimensional system $S_d$.
We shall denote by $M_d({\bf C})$
the algebra of $d\times d$ complex matrices
and by ${\cal S}_d$ the space of states (density matrices) of $S_d$, 
that is the convex set of positive $\rho\in M_d({\bf C})$ of unit trace.

Any hermiticity and trace-preserving linear map 
$\Lambda:{\cal S}_d\mapsto{\cal S}_d$ can be written as \cite{GKS}
\begin{equation}
\label{pos1}
\rho\mapsto\Lambda[\rho]=\sum_{k,i=0}^{d^2-1}\,
\lambda_{ki}\,F_k\,\rho\,F^\dagger_i\ ,
\end{equation}
where the $F_k$, $k=1,\ldots,d^2-1$, are traceless $d\times d$ matrices, 
forming together with the normalized identity $F_0:={\bf 1}_d/\sqrt{d}$ an orthonormal set in
$M_d({\bf C})$: ${\rm Tr}\bigl(F^\dagger_i F_k\bigr)=\delta_{ki}$,
$i,k=0,1,\ldots, d^2-1$;
the coefficients $\lambda_{ki}$ form a generic hermitian matrix such that
$\sum_{k,i=0}^{d^2-1}\lambda_{ki}F_i^\dagger F_k=1$.

Any linear map $\Lambda$ that is used to describe a physical state
transformation, must preserve the positivity of all states $\rho$,
otherwise the appearance of negative eigenvalues in $\Lambda[\rho]$
would spoil its statistical interpretation, which is based on the use
of its eigenvalues as probabilities.

The property of preserving the positivity of the spectrum of all
$\rho$ is called positivity; however, it is not sufficient to make $\Lambda$
fully physically consistent.
Indeed, the system $S_d$ may always be thought to be statistically
coupled to an ancilla $n$-level system $S_n$; one is thus forced to
consider the action ${\rm id}_n\otimes\Lambda$ over the compound
system $S_n+S_d$, where ${\rm id}_n$ is the identity action on the
ancilla.
In order to be fully physically consistent as a
state transformation, not only $\Lambda$ should be positive, but
also  ${\rm id}_n\otimes\Lambda$ for all $n$; 
such a property is called complete positivity \cite{CH1,Tak,Kra}.

Complete positivity is necessary because
of the existence of entangled states of the compound system $S_n+S_d$,
namely of states that cannot be written as factorized linear convex 
combinations
\begin{equation}
\label{sepst}
\rho^{\rm sep}_{S_n+S_d}=\sum_{ij}\,c_{ij}\,\rho_{S_n}^i\otimes\rho_{S_d}^j\ ,\quad
c_{ij}>0\ ,\quad\sum_{ij}c_{ij}=1\ .
\end{equation}
If only separable states as in (\ref{sepst}) existed,
then positivity of $\Lambda$ would be enough.
 
We briefly collect below some results concerning positivity and complete
positivity that will be used in the following.
\smallskip

In the space of states ${\cal S}_{d\times d}$ of the bipartite system $S_d+S_d$,
let us introduce the symmetric state
\begin{equation}
\label{pos2}
\vert\Psi^d_+\rangle=\frac{1}{\sqrt{d}}\sum_{j=1}^d\vert j\rangle\otimes\vert
j\rangle\ ,
\end{equation}
where $\vert j\rangle$, $j=1,2,\ldots,d$ is any fixed orthonormal
basis in ${\bf C}^d$,
and the corresponding projection
$P^d_+\equiv|\Psi^d_+\rangle\langle \Psi^d_+|$ $\in{\cal S}_{d\times d}$
onto it. 
\begin{theorem}{Theorem} \label{TH1}
The following results hold:
\begin{itemize} 
\item[(i)]
$\Lambda$ is positive on ${\cal S}_d$ if and only if
\begin{equation}
\label{pos}
\langle\phi\otimes\psi\vert\,
\big(\operatorname{id}_d\otimes\Lambda\big)[P^d_+]\,\vert\phi\otimes\psi\rangle=\frac{1}{d}
\langle\psi\vert\Lambda[\,\vert\phi^*\rangle\langle\phi^*\vert\,]
\vert\psi\rangle\geq 0
\end{equation}
for all normalized $\vert\phi\rangle$ and $\vert\psi\rangle\in{\bf C}^d$, with
$\vert\phi^*\rangle$ denoting the conjugate of $\vert\phi\rangle$ with
respect to the fixed orthonormal basis in ${\bf C}^d$ \cite{Jam,K}.
\item[(ii)]
$\Lambda$ is completely positive on ${\cal S}_d$ if and only if
\cite{CH1,HH}
\begin{equation}
\label{cpos}
\big(\operatorname{id}_d\otimes\Lambda\big)[P^d_+]\geq0\ .
\end{equation}
\item[(iii)]
A linear map $\Lambda$ is completely positive on ${\cal S}_d$
if and only if it can be expressed in the Kraus-Stinespring 
form~\cite{Tak,Kra} 
\begin{equation}
\label{cpos1}
\Lambda[\rho]=\sum_{j=0}^{d^2-1}\, G_j\,\rho\,G^\dagger_j\ ,
\end{equation} 
where the matrices $G_j$ need not be traceless and must satisfy the condition
$\sum_jG^\dagger_jG_j=1$ if one asks for trace-preservation.
\end{itemize}
\end{theorem}
\begin{definition}{Remark} \label{RM1}
Physically speaking, this Theorem states that, when 
$P\in{\cal S}_{d\times d}$ is a
generic one-dimensional projection, positivity of
$\Lambda$ guarantees the positivity of
$\big(\operatorname{id}_d\otimes\Lambda\big)[P]$ only when $P$ is separable, {\it i.e.}
$P=\vert\psi\rangle\langle\psi\vert\otimes\vert\phi\rangle\langle\phi\vert$.
Local operations must transform states of the bipartite system $S_d+S_d$ into
states keeping the positivity of the associated density matrices; 
therefore, the existence of entangled states as $P_+^d$ excludes that 
$\Lambda$, when only positive, may correspond to a physically
consistent local state-transformation.
\end{definition}
\begin{definition}{Remark} \label{RM2}
The positivity or complete positivity of $\Lambda$ depends
on the properties of the $d^2\times d^2$ matrix of coefficients $[\lambda_{ki}]$
in \eqref{pos1}.
Indeed, the Kraus-Stinespring form of completely positive maps corresponds to 
$[\lambda_{ki}]={\rm diag}(1,1,\ldots,1)$.
Vice versa, if  the matrix
$[\lambda_{ki}]$ is positive, diagonalizing 
$\lambda_{ki}=\sum_{j=0}^{d^2-1}\ell_j
U_{kj}U_{ij}^*$ and setting
$G_j=\sum_{k=0}^{d^2-1}\sqrt{\ell_k}U_{kj}F_k$, one recovers 
(\ref{cpos1}).
\end{definition}
\begin{definition}{Remark} \label{RM3}
The symmetric state $|\Psi_+^d\rangle$ and the corresponding
density matrix $P_+^d$ satisfy the following properties:
\begin{itemize}
\item[(i)] for all matrices $A$, $B$ acting on ${\bf C}^d$ one has
\begin{equation}
\label{prop1}
A\otimes B\vert\Psi^d_+\rangle={\bf 1}_d\otimes BA^T\vert\Psi^d_+\rangle=
AB^T\otimes{\bf 1}_d\vert\Psi^d_+\rangle\ ,
\end{equation}
\indent
where $A^T$, $B^T$ denote the transposed of $A$, $B$;
\item[(ii)] under partial transposition, $P^d_+$ gives rise to the flip operator
\begin{equation}
\label{prop2}
V=\sum_{a,b=1}^d\vert a\rangle\langle b\vert\otimes
\vert b\rangle\langle a\vert\,=\,d\,\big(T_d\otimes\operatorname{id}_d\big)[P^d_+]\ ,
\end{equation}
which is such that 
\begin{equation}
\label{prop3}
V\vert\psi\otimes\phi\rangle=\vert\phi\otimes\psi\rangle\ ,\quad
V\big(A\otimes B\big)V=B\otimes A \ .
\end{equation}
\end{itemize}
\end{definition}

Unlike for completely positive maps, there is no general prescription
on $[\lambda_{ki}]$ 
ensuring that $\Lambda$ preserve the positivity of $\rho$.
For instance, if $[\lambda_{ki}]$ is not positive, then, by separating positive 
and negative eigenvalues, one sees that every
$\Lambda$ can be written as the difference of two completely positive
maps:
\begin{equation}
\label{pos3}
\Lambda[\rho]=\sum_{\ell_j\geq0}\ell_j\,G_j\,\rho\,G_j^\dagger\ -\
\sum_{\ell_j<0}|\ell_j|\,G_j\,\rho\,G_j^\dagger\ ,
\end{equation} 
where the matrices $G_j:=\sum_{k=0}^{d^2-1}U_{kj}F_k$,
like the $F_k$'s,
are orthogonal.
However, no general rule is known that may allow us to recognize the
positivity of $\Lambda$ by looking at the eigenvalues $\ell_k$ and at
the matrices $G_k$.
One has to content oneself with sufficient conditions 
as the one which follows, that assumes the existence of just one
negative eigenvalue.
\begin{theorem}{Proposition} \label{PR1}
Suppose that 
$\ell_k\geq 0$ for all $k\neq p$, while $\ell_p<0$, whence
(\ref{pos3}) reads
\begin{equation}
\label{pos4}
\Lambda[\rho]=\sum_{\ell_k\geq0}\ell_k\,G_k\,\rho\,G_k^\dagger\ -\
|\ell_p|\,G_p\,\rho\,G_p^\dagger\ .
\end{equation}
If $\|G_p\|^2=M<1$ and $\ell_k\geq\frac{M}{1-M}|\ell_p|$, $k\neq p$, then $\Lambda$ is positive.
\end{theorem}
\noindent
The condition (\ref{pos}) is satisfied since
\begin{align*}
\langle\psi\vert
\Bigg(\sum_{k=0}^{d^2-1}
\ell_k\,G_k\vert\phi^*\rangle\langle\phi^*\vert G_k^\dagger\Bigg)
\vert\psi\rangle
&\geq\vert\ell_p\vert \Bigg(\frac{M}{1-M}\sum_{k\neq p}\left|
\langle\psi\vert G_k\vert\phi^*\rangle\right|^2
-\left|\langle\psi\vert G_p\vert\phi^*\rangle\right|^2\Bigg)\\
&=|\ell_p|\ \frac{M}{1-M}\Bigg(1\,-\,\frac{1}{M}\,\left|
\langle\psi\vert G_p\vert\phi^*\rangle
\right|^2\Bigg)\geq0\ ,
\end{align*}
where the last equality holds because the 
$G_k$'s form a basis in $M_d({\bf C})$, whence 
$$
\sum_{k=0}^{d^2-1}\left|\langle\psi\vert G_k\vert\phi^*\rangle\right|^2
=\operatorname{Tr}\Bigg[\Bigg(\sum_{k=0}^{d^2-1}G_k\operatorname{Tr}\Big[G_k^\dagger\vert\psi
\rangle\langle\phi^*\vert\Big]\Bigg)
\vert\phi^*\rangle\langle\psi\vert\Bigg]
=\langle\phi^*\vert\phi^*\rangle\langle\psi\vert\psi\rangle=1\ .
$$
{\hfill\rule{5pt}{5pt}}

\begin{definition}{Example}
Let $d=2$ and $\sigma_\alpha$, $\alpha=0,1,2,3$, 
be the Pauli matrices and the $2\times 2$ 
identity matrix $\sigma_0$.
Let $S_\alpha:{\cal S}_2\mapsto{\cal S}_2$ denote the positive map
$\rho\mapsto\sigma_\alpha \rho\sigma_\alpha$, and consider the maps
\begin{equation}
\label{TraceTansp}
\rho\mapsto\operatorname{Tr}_2[\rho]:=\frac{1}{2}\sum_{\alpha=0}^3S_\alpha[\rho]\ , \quad
\rho\mapsto T_2[\rho]:=\frac{1}{2}\sum_{\alpha=0}^3\varepsilon_\alpha
S_\alpha[\rho]\ ,
\end{equation}
where $\varepsilon_\alpha=1$ when $\alpha\neq 2$, whereas $\varepsilon_2=-1$.
By developing $\rho=\sum_{\alpha=0}^3\rho_\alpha\,\sigma_\alpha$, it is
straightforward to check that the first map amounts to taking the
trace and multiplying it by the identity $\sigma_0$;
it is completely positive: indeed,
by taking $F_\alpha=\sigma_\alpha/\sqrt{2}$ in \eqref{pos1}, one finds
$[\lambda_{\alpha\beta}]=\hbox{diag}(1,1,1,1)$. 
On the other hand, $T_2$ corresponds to transposition and is only positive, 
for $[\lambda_{\alpha\beta}]=\hbox{diag}(1,1,-1,1)$ and $\|F_\alpha\|^2=1/2$.
\end{definition}

As observed in Remark 2, a positive map $\Lambda$ that is not completely positive is unphysical because its ``extension'' by the identity map, \emph{i.e.} ${\rm id}_d\otimes\Lambda$, moves
entangled states as $P^+_d$ 
out of the space of states. However, exactly because
of this it may be used to detect entanglement \cite{HHH1}.
\begin{theorem}{Theorem}
A state $\rho\in{\cal S}_{d\times d}$ is entangled iff 
\begin{equation}
\label{pos6}
D_\Lambda[\rho]:=\operatorname{Tr}\Bigl[\big(\operatorname{id}_d\otimes\Lambda\big)[P^d_+]\,\rho\Bigr]\,<\,0
\end{equation}
for some positive map $\Lambda$ on ${\cal S}_d$.
\end{theorem}
Indeed, notice that according to \eqref{pos}, $D_\Lambda[\rho^{\rm sep}]\geq0$
for any separable state $\rho^{\rm sep}$.

\begin{definition}{Remark} 
For any map $\Lambda$, its dual $\Lambda^*$ 
is defined by ${\rm Tr}(\Lambda[X]\rho)={\rm Tr}(X\Lambda^*[\rho])$, 
for any $\rho$, $X$.
Then, the above result can be formulated by saying that $\rho\in{\cal
S}_{d\times d}$ is entangled if and only if there exists a positive map
$\Lambda$ on ${\cal S}_d$ such that $\big(\operatorname{id}_d\otimes\Lambda^*\big)[\rho]$ is
non-positive.
\end{definition}
As mentioned in the Introduction, an important example of positive, but not
completely positive map is the transposition $T_d$ 
(w.r.t. a fixed basis in ${\bf C}^d$);
through it, one can construct a subset of positive maps, the 
cone of decomposable maps.
\begin{definition}{Definition}
A positive map $\Lambda$ on ${\cal S}_d$ is decomposable if
\begin{equation}
\label{pos5}
\Lambda=\Gamma_1\,+\,\Gamma_2\circ T_d\ ,
\end{equation}
with $\Gamma_{1,2}$ completely positive on ${\cal S}_d$.
\end{definition}
\begin{definition}{Remark}
If $\big(\operatorname{id}_d\otimes T_d\big)[\rho]$ is positive, then 
$\big(\operatorname{id}_d\otimes \Lambda\big)[\rho]$ is also positive for any
decomposable $\Lambda$. Thus, if $\rho\in{\cal S}_{d\times d}$ is entangled,
but with positive partial transpose (PPT), a decomposable $\Lambda$
can not detect its entanglement since $D_\Lambda(\rho)\geq 0$.
\end{definition}
As a consequence, one has the following result.
\begin{theorem}{Proposition}
If $\Lambda$ is positive on ${\cal S}_d$,  
$\rho\in{\cal S}_{d\times d}$ is PPT and $D_\Lambda(\rho)<0$, then
$\Lambda$ is not decomposable and $\rho$ is entangled.
\end{theorem}
For $d=2$, 
all positive maps result decomposable \cite{Wor},
whence the transposition detects all the entangled states.
In other words,  $\big(\operatorname{id}_2\otimes T_2\big)[\rho]$ is
non-positive if and only if $\rho$ is entangled.
On the contrary, 
when $d\geq 3$, there are PPT states which are entangled
\cite{HHH0,CH2,Sto,H1}; their entanglement can not be
distilled by means of local operations and classical communication \cite{HHH2}, 
and it is called bound entanglement.
It is thus important not only to identify entanglement when present,
but also to qualify whether it is bound or not.

\subsection{Positive and completely positive semigroups}

The unitary time evolution for a system $S_d$
generated by a Hamiltonian operator results
automatically completely positive due to the criterion (iii)
in Theorem 1.
On the other hand, when the system is in contact with an external environment,
its subdynamics is in general not unitary, incorporating noise and
dissipative effects. In many physically interesting cases, such
dynamics can be approximated by families of
linear maps $\gamma_t$, $t\geq 0$ on ${\cal S}_d$ that are not invertible and
obey a semigroup composition
law, $\gamma_t\circ\gamma_s=\gamma_{t+s}$ \cite{Spo,AL,BP,BF0}.

Under the mild assumptions of continuity in time,
$\lim_{t\to0}\gamma_t=\operatorname{id}_d$, and preservation of hermiticity and trace, 
the semigroup has the form $\gamma_t=\exp(tL)$ with generator~\cite{GKS} 
\begin{equation}
\label{semg0}
L[\rho]=
-i\Bigl[H\,,\,\rho_t\Bigr]\,+\,\sum_{i,j=1}^{d^2-1}\,
C_{ij}\Bigl(F_i\rho_t
F^\dagger_j\,-\,\frac{1}{2}\Bigl\{F_j^\dagger F_i\,,\,\rho_t\Bigr\}\Bigr)\ ,
\end{equation}
where the $F_j$ are as in (\ref{pos1}), while the
matrix of coefficients $[C_{ij}]$, called the Kossakowski matrix, is hermitian.

Regarding the generated maps $\exp(tL)$, there are no general results on the form of
the Kossakowski matrix $[C_{ij}]$ such that they are
positive on ${\cal S}_d$, 
whereas they are completely positive if and only if $[C_{ij}]\geq 0$.

In the following we shall be interested in semigroups 
$\Gamma_t=\gamma_t^1\otimes\gamma_t^2$ on ${\cal S}_{d\times d}$ 
that are tensor products of semigroups $\gamma_t^{1,2}=\exp(tL_{1,2})$ 
on ${\cal S}_{d}$. 
Like in Proposition 5, one can give simple conditions on the 
Kossakowski matrices $C_{1,2}$ in the generators $L_{1,2}$ that are 
sufficient for the positivity of $\Gamma_t$ \cite{BFP2}. 
\begin{theorem}{Proposition}
Suppose that the non-Hamiltonian terms in the generators of 
$\gamma^{1,2}_t$ are as follows,
\begin{equation}
\label{gen3}
L_{i}[\rho]=\sum_{\ell=1}^{d^2-1}c_{i}^\ell\Bigl(G_\ell^{i}\,
\rho\,G_\ell^{i}\,-\,\frac{1}{2}\Bigl\{(G_\ell^{i})^2\,,\,\rho\Bigr\}
\Bigr)\ ,\quad c^\ell_{i}\in{\bf R}\ ,
\end{equation}
for $i=1,2$, where $G^{i}_\ell\in M_{d}({\bf C})$, together with 
$G^{i}_0={\bf 1}_2/\sqrt{d}$, constitute two 
orthonormal sets of hermitian matrices.
Suppose that $c_1^\ell>0$ for all $\ell=1,2,\ldots,d^2-1$, and that 
$c_2^k=-|c_2^k|<0$, for one index $k$, while $c_2^\ell>0$ when
$\ell\neq k$; then, the map
$\Gamma_t=\gamma^1_t \otimes\gamma^2_t$ is positive if
$c_1^\ell\geq |c_2^k|$, $\ell=1,2,\ldots,d^2-1$ and
$c_2^\ell\geq |c_2^k|$, $\ell=1,2,\ldots,d^2-1,\,\ell\neq k$. 
\end{theorem}
\begin{definition}{Example}
Let $d=2$, set $H=0$, $F_i=\sigma_i/\sqrt{2}$ and
choose as Kossakowski matrices
\begin{equation}
\label{semg1}
C_1=\begin{pmatrix}1&0&0\cr0&1&0\cr0&0&1\end{pmatrix}\ ,\qquad
C_2=\begin{pmatrix}1&0&0\cr0&-1&0\cr0&0&1\end{pmatrix}\ .
\end{equation}
Using the notation of Example 6, the generators are given by
\begin{equation}
\label{semg2}
\partial_t\gamma_t^1[\rho]=L_1[\rho]:=\frac{1}{2}\Bigl(\sum_{i=1}^3
S_i[\rho]\,-\,3\,\rho\Bigr)\ ,\quad
\partial_t\gamma_t^2[\rho]=L_2[\rho]:=\frac{1}{2}\Bigl(\sum_{i=1}^3
\varepsilon_i\,S_i[\rho]\,-\,\,\rho\Bigr)\ ,
\end{equation}
yielding
\begin{equation}
\label{semg3}
\gamma^1_t={\rm e}^{-2t}\operatorname{id}_2+\frac{1-{\rm e}^{-2t}}{2}\operatorname{Tr}_2\ ,\quad
\gamma_t^2=\frac{1+{\rm e}^{-2t}}{2}\operatorname{id}_2+\frac{1-{\rm e}^{-2t}}{2}T_2\ .
\end{equation}
As $T_4=T_2\otimes T_2$ and $\operatorname{Tr}_2\circ T_2=\operatorname{Tr}_2$, 
$\Gamma_t=\gamma_t^1\otimes\gamma^2_t$ can be written as
\begin{eqnarray}
\nonumber
\Gamma_t&=&{\rm e}^{-2t}\frac{1+{\rm e}^{-2t}}{2}\operatorname{id}_4\,+\,
\frac{1-{\rm e}^{-4t}}{4}\operatorname{Tr}_2\otimes\operatorname{id}_2\\
\label{semg4}
&+&\frac{1-{\rm e}^{-2t}}{2}\Bigl(
{\rm e}^{-2t}T_2\otimes\operatorname{id}_2\,+\,\frac{1-{\rm
e}^{-2t}}{2}\operatorname{Tr}_2\otimes\operatorname{id}_2\Bigr)\circ T_4\ .
\end{eqnarray}
The semigroup $\Gamma_t$ is written as
$\Gamma_t=\Gamma^1_t+\Gamma^2_t\circ T_4$, with $\Gamma^1_t$ completely
positive for all $t\geq0$, while $\Gamma_t^2$ is such only for $t=0$ and
$t\geq t^*:=(\log3)/2$.
It follows that $\Gamma_t$ is surely decomposable for $t=0$
($\Gamma_0=\operatorname{id}_{16}$) and for $t\geq t^*$; on the other hand, as shown below
the positive maps $\Gamma_t$ result indecomposable for
$0<t<t^*$.
\end{definition}

\section{Lattice states}

Let $L_{16}$ be the square lattice 
$\big\{(\alpha,\beta):\alpha\,,\,\beta=0,1,2,3\big\}$ with $16$
elements, that we can split into columns 
${\cal C}_\alpha=\big\{(\alpha,\beta):\beta=0,1,2,3\big\}$
and rows ${\cal R}_\beta=\big\{(\alpha,\beta)\,:\alpha=0,1,2,3\big\}$.
To each site $(\alpha,\beta)$ we will
associate the following $4\times 4$ tensor products of Pauli matrices
$\sigma_{\alpha\beta}:=\sigma_\alpha\otimes\sigma_\beta$.
By acting with ${\bf 1}_4\otimes\sigma_{\alpha\beta}$ 
on the symmetric state $|\Psi_+^4\rangle$
one constructs an orthonormal basis of maximally entangled
vectors in ${\bf C}^{16}$,
\begin{equation}
\label{symmvec}
|\Psi_{\alpha\beta}\rangle:=\big({\bf 1}_4\otimes\sigma_{\alpha\beta}\big)\, |\Psi_+^4\rangle,
\end{equation}
with the corresponding orthogonal projections, 
\begin{equation}
\label{sqlatst1}
P_{\alpha\beta}:=|\Psi_{\alpha\beta}\rangle\langle\Psi_{\alpha\beta}|=
\big({\bf 1}_4\otimes\sigma_{\alpha\beta}\big)\,P^4_+\,
\big({\bf 1}_4\otimes\sigma_{\alpha\beta}\big)\ ,\qquad 
P_{\alpha\beta}\,P_{\gamma\varepsilon}=\,
\delta_{\alpha\gamma}\,\delta_{\beta\varepsilon}\ P_{\alpha\beta}\ .
\end{equation}

We will focus on states in ${\cal S}_4\times{\cal S}_4$ that are diagonal with respect
to the just introduced basis $\big\{|\Psi_{\alpha\beta}\rangle\big\}$,
\begin{equation}
\label{statepi}
\rho_\pi:=\sum_{(\alpha,\beta)\in L_{16}} \pi_{\alpha\beta}\, \,P_{\alpha\beta}
\ ,\quad \pi_{\alpha\beta}\geq0\ ,\ \sum_{(\alpha\beta)\in
L_{16}}\pi_{\alpha\beta}=1\ .
\end{equation}
Actually, for sake of simplicity we shall further limit the discussion to those
states $\rho_I$ for which all the non-vanishing weights 
$\pi_{\alpha\beta}$ are equal.
They are completely characterized by the subset $I=\{(\alpha,\beta):
\pi_{\alpha\beta}\neq0\}\subseteq L_{16}$, and take the form
\begin{equation}
\label{eqsqlatst}
\rho_I=\frac{1}{N_I}\sum_{(\alpha,\beta)\in I}\, P_{\alpha\beta}\ ,
\end{equation}
where $N_I$ is the number of elements of $I$.

We shall now characterize all states $\rho_I$ that are PPT and
study their entanglement properties.

\subsection{PPT States}

We first act with the transposition
on the first subsystem in \eqref{statepi}: 
\begin{equation}
\label{ptr1}
\rho_\pi^{T_1}:=\big(T_4\otimes{\rm id}_4\big)[\rho_\pi]=
\frac{1}{4}\,\sum_{(\alpha,\beta)\in L_{16}}\pi_{\alpha\beta}\,V_{\alpha\beta}\ ,\qquad
V_{\alpha\beta}:=\big({\bf 1}_4\otimes\sigma_{\alpha\beta}\big)
\,V\,\big({\bf 1}_4\otimes\sigma_{\alpha\beta}\big)\ .
\end{equation}

\begin{theorem}{Lemma}
\label{LM1}
The matrices $V_{\alpha\beta}$ are self-adjoint 
and their spectral decomposition is as follows: 
$\displaystyle
V_{\alpha\beta}=\sum_{(\gamma,\delta)\in
L_{16}}\xi_{\alpha\gamma}\,\xi_{\beta\delta}\, P_{\gamma\delta}$\quad where\quad
$\xi_{\alpha\gamma}=1\,-\,2\,\delta_{|\alpha-\gamma|,2}$ .
\end{theorem}
From (\ref{prop3}) and $V\vert\Psi^4_+\rangle=\vert\Psi_+^4\rangle$, 
it follows that
\begin{eqnarray*}
V_{\alpha\beta}\vert\Psi_{\gamma\delta}\rangle&=&
\varepsilon_\alpha\varepsilon_\gamma\varepsilon_\beta\varepsilon_\delta
\Big\{ {\bf 1}_4\otimes\Big[\big(\sigma_{\alpha}\sigma_\gamma\sigma_\alpha\big)
\otimes\big(\sigma_\beta\sigma_\delta\sigma_\beta\big)\Big]\Big\}
\vert\Psi_+^4\rangle\\
&=&
\big(\varepsilon_\alpha\varepsilon_\gamma\eta_{\alpha\gamma}\big)\
\big(\varepsilon_\beta\varepsilon_\delta\eta_{\beta\delta}\big)
\ \vert\Psi_{\gamma\delta}\rangle\ ,
\end{eqnarray*}
where $\eta_{\alpha\gamma}$ takes the values $\pm 1$
according to the table
$$
\hbox{\begin{tabular}{l||r|r|r|r}
$\gamma\backslash\alpha$&0&1&2&3\\ \hline\hline
0&1&1&1&1\\ \hline
1&1&1&-1&-1\\ \hline
2&1&-1&1&-1\\ \hline
3&1&-1&-1&1\\ 
\end{tabular}}
$$
and $\varepsilon_\alpha$ has been defined in Example 6.
Setting 
$\xi_{\alpha\gamma}=\varepsilon_\alpha\varepsilon_\gamma\,\eta_{\alpha\gamma}$,
the result follows by direct inspection.
\hfill\rule{5pt}{5pt}

\begin{theorem}{Proposition}
\label{PR3}
A state $\rho_I$ is PPT if and only if for any lattice point $(\alpha,\beta)\in L_{16}$
the corresponding column ${\cal C}_\alpha$ and row ${\cal R}_\beta$  do not contain
more than $N_I/2$ elements of $I$, the point $(\alpha,\beta)$ itself excluded.
\end{theorem}
It proves convenient to consider the convex set of states
$\rho_\pi$ introduced in \eqref{statepi}.
Recalling the explicit form of the coefficients $\xi_{\alpha\gamma}$ above, 
we introduce the bijection
$(\alpha,\beta)\mapsto(\widetilde{\alpha},\widetilde{\beta})$, where
$\widetilde{\mu}:=(\mu+2)\!\!\mod\!(4)$, $\mu=0,\ldots,3$.
Then, from (\ref{ptr1}), using  
the previous Lemma, we obtain the spectral decomposition
\begin{eqnarray}
\label{ppt1}
\rho_\pi^{T_1}&=&\sum_{(\gamma,\delta)\in
L_{16}}\,
\frac{1}{4}\Bigl[1-2\Bigl(Q^{(1)}_{\widetilde{\gamma}\widetilde{\delta}}+
Q^{(2)}_{\widetilde{\gamma}\widetilde{\delta}}\Bigr)\Bigr]\, P_{\gamma\delta}\\
\label{ppt2}
Q^{(1)}_{\widetilde{\gamma}\widetilde{\delta}}&:=&
\sum_{\beta\neq\widetilde{\delta}}\pi_{\widetilde{\gamma}\beta}\ ,\qquad
Q^{(2)}_{\widetilde{\gamma}\widetilde{\delta}}:=
\sum_{\alpha\neq\widetilde{\gamma}}\pi_{\alpha\widetilde{\delta}}\ ,
\end{eqnarray}
whence $\rho_\pi$ is PPT if and only if 
$\displaystyle
Q^{(1)}_{\widetilde{\gamma}\widetilde{\delta}}+
Q^{(2)}_{\widetilde{\gamma}\widetilde{\delta}}\leq\frac{1}{2}$ for all
$(\gamma,\delta)\in L_{16}$.
Because of the bijection
$(\mu,\nu)\mapsto(\widetilde{\mu},\widetilde{\nu})$, 
one has, equivalently, that $\rho_\pi$ is PPT if and only if 
$\displaystyle Q^{(1)}_{\gamma\delta}+
Q^{(2)}_{\gamma\delta}\leq\frac{1}{2}$ for all
$(\gamma,\delta)\in L_{16}$.
Setting $\pi_{\alpha\beta}=1/N_I$ on $I\subseteq L_{16}$ and
$\pi_{\alpha\beta}=0$ otherwise, the result follows, for
the quantities $Q^{(1,2)}$
just count how many contributions come from any given row 
and column of the lattice, the intersection point $(\alpha,\beta)$ excluded.
\hfill\rule{5pt}{5pt}


\begin{definition}{Remark} \label{ordering}
In order to concretely construct the states $\rho_I$ that are PPT,
one can proceed by first selecting those subsets $I\subseteq L_{16}$
containing a decreasing number of elements in the ordered
columns ${\cal C}_0$, ${\cal C}_1$, ${\cal C}_2$, ${\cal C}_3$,
with those in ${\cal C}_0$ piled one over the other, without holes, and
further satisfying the hypothesis of Proposition 15.
\end{definition}

\begin{definition}{Example} \label{exstates}
Representing the elements of $I$ by crosses inserted at the corresponding
sites of the lattice $L_{16}$, one can associate to each state $\rho_I$
a graph. Using this correspondence, we give below some examples PPT states.
\bigskip
$$
N_I=4\, :\quad
\hbox{\begin{tabular}{r|r|r|r|r}
3&&&&$\times$\\ \hline
2&&&$\times$&\\ \hline
1&&$\times$&&\\ \hline
0&$\times$&&&\\ \hline
&0&1&2&3\\ 
\end{tabular}}\qquad \qquad
\hbox{\begin{tabular}{r|r|r|r|r}
3&&&$\phantom{\times}$&$\phantom{\times}$\\ \hline
2&&&&\\ \hline
1&$\times$&$\times$&&\\ \hline
0&$\times$&$\times$&&\\ \hline
&0&1&2&3\\
\end{tabular}}
$$
\smallskip

$$
N_I=6\, :\quad
\hbox{\begin{tabular}{r|r|r|r|r}
3&&&&$\phantom{\times}$\\ \hline
2&&&&\\ \hline
1&$\times$&$\times$&$\times$&\\ \hline
0&$\times$&$\times$&$\times$&\\ \hline
&0&1&2&3\\ 
\end{tabular}}\qquad \qquad
\hbox{\begin{tabular}{r|r|r|r|r}
3&&&&\\ \hline
2&$\times$&&&$\times$\\ \hline
1&$\times$&$\times$&&\\ \hline
0&$\times$&&$\times$&\\ \hline
&0&1&2&3\\
\end{tabular}}
$$
\smallskip

$$
N_I=8\, :\quad
\hbox{\begin{tabular}{r|r|r|r|r}
3&&&&$\phantom{\times}$\\ \hline
2&$\times$&$\times$&$\times$&\\ \hline
1&$\times$&$\times$&&\\ \hline
0&$\times$&$\times$&$\times$&\\ \hline
&0&1&2&3\\ 
\end{tabular}}\qquad \qquad
\hbox{\begin{tabular}{r|r|r|r|r}
3&&&&\\ \hline
2&$\times$&$\times$&&$\times$\\ \hline
1&$\times$&$\times$&&\\ \hline
0&$\times$&$\times$&$\times$&\\ \hline
&0&1&2&3\\
\end{tabular}}
$$
\smallskip

$$
N_I=10\ :\quad
\hbox{\begin{tabular}{r|r|r|r|r}
3&&&&\\ \hline
2&$\times$&$\times$&$\times$&$\times$\\ \hline
1&$\times$&$\times$&&$\times$\\ \hline
0&$\times$&$\times$&$\times$&\\ \hline
&0&1&2&3\\
\end{tabular}}\qquad \qquad
\hbox{\begin{tabular}{r|r|r|r|r}
3&&&&\\ \hline
2&$\times$&$\times$&$\times$&$\times$\\ \hline
1&$\times$&$\times$&$\times$&\\ \hline
0&$\times$&$\times$&$\times$&\\ \hline
&0&1&2&3\\
\end{tabular}}
$$
\noindent
In addition, it is instructive to give some examples of states whose partial transpose is not positive; in each of them the condition of Proposition 15 is not satisfied, as easily seen by considering the highlighted columns and rows.
\bigskip
$$
N_I=4\, :\quad
\hbox{\begin{tabular}{r|r|>{\columncolor{gray}}r|r|r}
3&&&$\phantom{\times}$&$\phantom{\times}$\\ \hline
2&&$\times$&&\\ \hline
\rowcolor{gray}$1$&$\times$&&&\\ \hline
0&$\times$&$\times$&&\\ \hline
&0&$1$&2&3\\
\end{tabular}}
\qquad \qquad
N_I=5\,:\quad
\hbox{\begin{tabular}{r|r|r|>{\columncolor{gray}}r|r}
3&&&&$\phantom{\times}$\\ \hline
2&&&&\\ \hline
\rowcolor{gray}1&$\times$&$\times$&&\\ \hline
0&$\times$&$\times$&$\times$&\\ \hline
&0&1&2&3\\
\end{tabular}}
$$
\smallskip

$$
N_I=6\, :\quad
\hbox{\begin{tabular}{r|r|r|>{\columncolor{gray}}r|r}
3&&&&$\phantom{\times}$\\ \hline
2&&&$\times$&\\ \hline
\rowcolor{gray}1&$\times$&$\times$&&\\ \hline
0&$\times$&$\times$&$\times$&\\ \hline
&0&1&2&3\\ 
\end{tabular}}\qquad \qquad
N_I=7\,:\quad
\hbox{\begin{tabular}{r|r|r|r|>{\columncolor{gray}}r}
3&&&&\\ \hline
2&&&&\\ \hline
\rowcolor{gray}1&$\times$&$\times$&$\times$&\\ \hline
0&$\times$&$\times$&$\times$&$\times$\\ \hline
&0&1&2&3\\
\end{tabular}}
$$
\smallskip
\end{definition}

All lattice states $\rho_I$ that result PPT can be obtained by those
defined in Remark 16 by applying suitable local unitary transformations.
Indeed, one has:

\begin{theorem}{Proposition}
Let $U$ and $W$ be $4\times 4$ unitary matrices, transforming
$\sigma_{\alpha\beta}\mapsto W\sigma_{\alpha\beta}U^\dagger$ 
into another element $\sigma_{\gamma\delta}$ up to a phase.
Then, any PPT state $\rho_I$ is mapped into another PPT
state $\rho_I$ of the same rank,
\begin{equation}
\label{eqrel}
\rho_{\hat I}:=\big(U^*\otimes W\big)\,\rho_I\,\big(U^T\otimes W^\dagger\big)\ ,
\end{equation}
where $\hat I\subseteq L_{16}$ is the image of $I$ under the
one-to-one correspondence $(\alpha,\beta)\mapsto(\gamma,\delta)$.
\end{theorem}
The result follows at once from the definition of $P_{\alpha\beta}$,
the property (\ref{prop1}) of completely symmetric states
and the hypothesis; in fact 
\begin{eqnarray*}
\big(U^*\otimes W\big)\,P_{\alpha\beta}\,\big(U^T\otimes W^\dagger\big)&=&
\Big[{\bf 1}_4\otimes \big(W\sigma_{\alpha\beta}U^\dagger\big)\Big]
\vert\Psi^4_+\rangle\langle\Psi^4_+\vert
\Big[{\bf 1}_4\otimes \big(U\sigma_{\alpha\beta}W^\dagger\big)\Big]\\
&=&P_{\gamma\delta} \ ,
\end{eqnarray*}
and the local unitary operations preserve the PPT property.
\hfill\rule{5pt}{5pt}

\medskip

By properly choosing $U$ and $W$, 
one can permute columns and rows.
Further, by taking $U$ and $W$ both equal to the flip operator $V$ in
(\ref{prop2}), one can go from any $\rho_I$ to the one with column and row
contributions exchanged.
More in general, one can  subdivide the PPT states $\rho_I$
into equivalence classes with respect to the relation given by (\ref{eqrel}). 

From the previous considerations, by direct inspection one can conclude that
for $N_I=1,2,3,5,7$, there are no states $\rho_I$ that are PPT, while they are
all PPT for $N_I\geq12$.

\subsection{Bound Entangled States}

Having fully characterized the states $\rho_I$ that are PPT, 
the next task is to determine
which of them are also entangled.
Unfortunately, we are not able to obtain a complete classification.
Nevertheless, by employing the semigroup of positive maps $\Gamma_t$
constructed in the previous section we can discuss
the entanglement properties of some non-trivial
classes of PPT states $\rho_I$. 

We first observe that: 

\vspace{-.3cm}

\begin{itemize}
\item[(i)] from general arguments rank-$4$ states cannot
be entangled~\cite{HLV};
\item[(ii)] all PPT lattice states $\rho_I$, with $N_I=15$, are separable
isotropic states; indeed, thanks to Proposition 18, all of them are in the
equivalence class of $\rho_{15}=(1-P_{00})/15$, with fidelity ${\cal F}=
\operatorname{Tr}[\rho_{15}\, P_{00}]=\,0$ ({\it cf.} \cite{HH}).
\end{itemize}

Further, we recall the following result discussed in \cite{BFP2}.

\begin{definition}{Example}
Recall the definition of the pairing $D_\Lambda$ in~(\ref{pos6})
and take for $\Lambda$ the map $\Gamma_t$ as in (\ref{semg4}); consider the
following state $\rho_I\in{\cal S}_4$, that according to the rule
in Proposition 15 is PPT
\begin{equation}
\label{sqltst1}
\rho_I=\frac{1}{6}\Bigl(P_{02}+P_{11}+P_{23}+P_{31}+P_{32}+P_{33}\Bigr)\ .
\end{equation}
It turns out that $D_{\Gamma_t}(\rho_I)<0$ for $0<t<(\ln3)/2$.  
As a consequence, $\rho$ is bound entangled and, in this range of
times, the positive maps $\Gamma_t$ turn out to be indecomposable.
\end{definition}
\medskip

Actually, the positive semigroup $\Gamma_t$ is able to detect the entanglement
of other PPT lattice states $\rho_I$. Instead of using the test (\ref{pos6}),
we will follow a different, equivalent method, based on Remark 8:
if for a PPT state $\rho_I$, any of the eigenvalues of
$\operatorname{id}_4\otimes\Gamma_t[\rho_I]$ is negative, then we can conclude
that $\rho_I$ is bound entangled (notice that the dual of $\Gamma_t$
coincides with $\Gamma_t$ itself).

It is straightforward to check that
\begin{equation}
\nonumber
\langle\Psi_{\gamma\delta}\vert\big(\operatorname{id}_4\otimes\Gamma_t\big)[P_{\alpha\beta}]
\vert\Psi_{\mu\nu}\rangle
=\delta_{\gamma\mu}\, \delta_{\delta\nu}\ p^{\mu\nu}_{\alpha\beta}(t)\ ,
\end{equation}
with
\begin{equation}
\label{PPTES2}
p^{\mu\nu}_{\alpha\beta}(t):=\frac{1}{16}\Bigl(
4{\rm e}^{-2t}\ \delta_{\alpha\mu}\ +\ 1\ -\ {\rm e}^{-2t}\Bigr)\
\times\
\Big[2\Bigl(1+{\rm e}^{-2t}\Bigr)\delta_{\beta\nu}\ +\
\Bigl(1-{\rm e}^{-2t}\Bigr)\xi_{\beta\nu}\Big]\ ,
\end{equation}
where $\xi_{\beta\delta}$ has been introduced in Lemma 14.
Therefore, we can write the following
spectral decomposition
\begin{equation}
\label{PPTES3}
\big(\operatorname{id}_4\otimes\Gamma_t\big)[\rho_I]=\sum_{(\mu,\nu)\in L_{16}}\,
R_{\mu\nu}(t)\ P_{\mu\nu}\ ,\quad 
R_{\mu\nu}(t):=\frac{1}{N_I}\sum_{(\alpha,\beta)\in
I}p^{\mu\nu}_{\alpha\beta}(t)\ .
\end{equation}
If one of the eigenvalues $R_{\mu\nu}(t)<0$, the PPT 
state $\rho_I$ is entangled and its entanglement of bound type and
thus undistillable.

\begin{theorem}{Proposition}
A PPT lattice state $\rho_I$ is entangled if there
exist a column ${\cal C}_\gamma$ and a row ${\cal R}_\delta$ of $L_{16}$
whose intersection with the subset $I$ contains only one element, $(\gamma,\delta')$,
$\delta'\neq\delta$, or
$(\gamma',\delta)$, $\gamma'\neq\gamma$.
\end{theorem}
By expanding the eigenvalues of $\big(\operatorname{id}_4\otimes\Gamma_t\big)[\rho_I]$ for small $t$,
one gets
\begin{equation}
\label{PPTES4}
p^{\mu\nu}_{\alpha\beta}(t)\simeq
(1-3t)\delta_{\alpha\mu}\delta_{\beta\nu}\ +\ 
\frac{t}{2}\,\Bigl(\delta_{\alpha\mu}\,\xi_{\beta\nu}\ +\
\delta_{\beta\nu}\Bigr)
\ .
\end{equation} 
Let us suppose that the only element of $I$ in the column
${\cal C}_\gamma$ and the row ${\cal R}_\delta$ be $(\gamma,\delta')$,
$\delta'\neq\delta$.
According to Proposition 18, we can always permute rows and
columns so that $\gamma=\delta=3$, and
$\delta'=1$, while remaining in the same equivalence class.
According to the assumptions and to the definition of
$\xi_{\nu_\beta}$ in Lemma 14, it turns out
that, for small positive $t$, the eigenvalue $R_{33}$ is negative:
\begin{equation}
R_{33}(t)=\frac{1}{N_I}\sum_{(\alpha,\beta)\in
I}p^{33}_{\alpha\beta}(t)\,=\,
-\frac{t}{2N_I}\ .
\end{equation}
as it can be easily checked by inserting \eqref{PPTES4} into \eqref{PPTES3}.
\hfill \rule{5pt}{5pt}

\medskip

Notice that the previous Proposition gives only a sufficient condition
for a state $\rho_I$ to be entangled and further that its
hypothesis can be satisfied only when $N_I\leq 10$.
Nevertheless, using it we are able to conclude that
among the PPT cases presented in Example 17, with $6\leq N_I\leq10$, 
those on the right result entangled. On the other hand,
Proposition 20 is inconclusive for what concerns the states
on the left; indeed, in those case,
$\big(\operatorname{id}_4\otimes\Lambda\big)[\rho_I]$
with $\Lambda=\Gamma_t$ has no negative eigenvalues for small $t$,
but these might occur for a different $\Lambda$.

Actually, we have been unable to find a bound entangled lattice state $\rho_I$
that is not detected by the positive map $\Gamma_t$; this fact
may suggest that $\Gamma_t$ is sufficient to classify all
entangled PPT state, within the studied class.

For instance, the rank-$6$ PPT state on the left in Example 17
results separable. In order to prove this,
first use Proposition 18 to transform it into the following element 
of its equivalence class:
$$
\rho_6=\frac{1}{6}\left(P_{00}+P_{01}+P_{02}+P_{30}+P_{31}+P_{32}\right)
\ .
$$
It is then lengthy, but not difficult to decompose
\begin{eqnarray*}
\rho_6&=&\frac{1}{12}\Bigl[P^{(1)}_{0+}+P^{(1)}_{0-}+P^{(1)}_{1+}+P^{(1)}_{1-}
+P^{(2)}_{0+}+P^{(2)}_{0-}+P^{(2)}_{1+}+P^{(2)}_{1-}\Bigr]\\
&+&\frac{1}{12}\Bigl[Q_3+Q_4+Q_5+Q_6\Bigr]\ ,
\end{eqnarray*}
where the $P$'s and $Q$'s are projections onto the separable states:

\begin{eqnarray*}
\vert\Psi^{(1)}_{0\pm}\rangle&=&
\frac{\vert00\rangle\pm\vert01\rangle}{\sqrt{2}}\otimes
\frac{\vert00\rangle\pm\vert01\rangle}{\sqrt{2}}\qquad
\vert\Psi^{(2)}_{0\pm}\rangle=
\frac{\vert00\rangle\pm i\vert01\rangle}{\sqrt{2}}\otimes
\frac{\vert00\rangle\mp i\vert01\rangle}{\sqrt{2}}\ ,
\\
\vert\Psi^{(1)}_{1\pm}\rangle&=&
\frac{\vert10\rangle\pm\vert11\rangle}{\sqrt{2}}\otimes
\frac{\vert10\rangle\pm\vert11\rangle}{\sqrt{2}}\qquad
\vert\Psi^{(2)}_{1\pm}\rangle=
\frac{\vert10\rangle\pm i\vert11\rangle}{\sqrt{2}}\otimes
\frac{\vert10\rangle\mp i\vert11\rangle}{\sqrt{2}}\ ,
\end{eqnarray*}
\begin{eqnarray*}
\vert\Phi_3\rangle&=&\vert00\rangle\otimes\vert01\rangle\hskip 3cm\ \
\vert\Phi_4\rangle=\vert01\rangle\otimes\vert00\rangle\ ,\\
\vert\Phi_5\rangle&=&\vert10\rangle\otimes\vert11\rangle\hskip 3cm\ \
\vert\Phi_6\rangle=\vert11\rangle\otimes\vert10\rangle\ ,
\end{eqnarray*}
respectively,
with $\vert0\rangle$ and $\vert1\rangle$ eigenstates of $\sigma_3$.

Another example is given by the already mentioned rank-15 states, but the
same conclusion holds also for rank-14 states. Indeed, all these latter
states are in a sigle equivalence class, whose representative
element is given by
$$
\begin{tabular}{c|c|c|c|c}
3&$\circ$&$\circ$&&\\ \hline
2&$\circ$&$\circ$&$\times$&$\times$\\ \hline
1&$\diamond$&$\diamond$&$\times$&$\times$\\ \hline
0&$\diamond$&$\diamond$&$\times$&$\times$\\ \hline
&0&1&2&3\\
\end{tabular}
$$
The diamonds, circles and crosses identify
lattice states of rank four and six that we have already
shown to be separable and whose linear convex combination
gives the chosen rank-14 lattice state.

We conclude by mentioning that most of the results presented here
can be generalized to the case of higher dimensional bipartite systems.
In particular, one can construct $2^N\times 2^N$-dimensional 
states whose bound entanglement is detected by dynamical semigroups 
of positive, non-decomposable maps that generalize the 
map $\Gamma_t$ discussed in Example 13 \cite{P}.


\end{document}